# Interface dynamics under nonequilibrium conditions: from a self-propelled droplet to dynamic pattern evolution

Yong-Jun Chen[1,3] and Kenichi Yoshikawa[1,2*]

[1]Department of Physics, Graduate School of Science, Kyoto University, Oiwake-cho, Kitashirakawa, Sakyo-ku, Kyoto 606-8502, Japan

[2]Spatio-Temporal Order Project, ICORP, JST (Japan Science and Technology Agency), Kyoto 606-8502, Japan

[3]Nanosystems Research Institute, National Institute of Advanced Industrial Science and Technology (AIST), 1-1-1 Higashi, Tsukuba, Ibaraki, 305-8565, Japan

Email: yongjunchen@ni.aist.go.jp (For Yong-Jun Chen)

Email: yoshikaw@scphys.kyoto-u.ac.jp (For Kenichi Yoshikawa)

* Author to whom correspondence should be addressed.

Abstract:

In this article, we describe the instability of a contact line under nonequilibrium conditions mainly based on the results of our recent studies. Two experimental examples are presented: the self-propelled motion of a liquid droplet and spontaneous dynamic pattern formation. For the self-propelled motion of a droplet, we introduce an experiment in which a droplet of aniline sitting on an aqueous layer moves spontaneously at an air-water interface. The spontaneous symmetry breaking of Marangoni-driven spreading causes regular motion. In a circular Petri dish, the droplet exhibits either beeline motion or circular motion. On the other hand, we show the emergence of a dynamic labyrinthine pattern caused by dewetting of a metastable thin film from the air-water interface. The contact line between the organic phase and aqueous phase forms a unique spatio-temporal pattern characterized as a dynamic labyrinthine. Motion of the contact line is controlled by diffusion processes. We propose a theoretical model to interpret essential aspects of the observed dynamic behavior.

## 1. Introduction

Interfaces are ubiquitous in daily life and in various scientific areas. For example, when a liquid is placed on a solid substrate, the interface between the liquid and solid will undergo transient motion, i. e., wetting or dewetting, depending on the surface condition. Interface dynamics play a role in many industrial processes [1]. This ubiquity has prompted numerous studies on interface dynamics. Interface dynamics is related to the nonequilibricity of the system, which might be induced by a chemical Marangoni effect [2-4], a potential gradient due to a difference in composition or density [5-7], a temperature gradient [8] and so on. A conservative system with a state that deviates from equilibrium favors motion, which is directed toward the equilibrium state of the system. However, for a dispersive system under thermodynamically open conditions, persistent motion is generated [4]. Thus, a spatio-temporal pattern emerges spontaneously. Such dispersive motion is well-known in living systems [9]. Living systems consume energy and show self-locomotion. Due to the importance of understanding biological dispersive systems and the practical implications of an industrial application, there is great interest in research on interface dynamics [1, 9]. Recently, we reported the spontaneous motion of a liquid droplet and a dynamic labyrinthine pattern in an active liquid film [10, 11]. Interface dynamics, which is induced by a Marangoni-driven spreading process, exhibits spatial-temporal structures. The symmetry breaking provides the driving force for interface motion. Dispersion in the vicinity of the interface supplies the kinetic energy for such interface dynamics.

## 2. Self-propelled motion of a droplet

When the balance of surface tension around a droplet is broken, a liquid droplet undergoes net motion [10]. Figure 1 shows an example of the self-propelled motion of an aniline droplet on an aqueous phase induced by Marangoni-driven spreading. A Petri dish is filled with aniline aqueous solution. An aniline droplet was placed on the air-solution interface. The droplet started to move spontaneously. The droplet exhibits either circular or beeline motion in a circular Petri dish, depending on the initial conditions. When the volume of the droplet is large enough, the droplet moves continuously for several hours. The wall apparently repels the droplet, as shown in the figure. The interaction between the droplet and the wall is subtle. A droplet in beeline motion prefers to spontaneously switch to circular motion over the long term after it bumps with the wall (Fig. 2). During circular motion, the distance between the droplet and the wall is within a certain range, as shown in Fig. 2(b). Figure 1(b) shows the spatio-temporal evolution of beeline motion between two parallel walls in a square vessel. The droplet undergoes back-and-forth motion between the two walls. The walls repel the droplet, which decelerates when it approaches the walls. The repulsive force from the walls, as evaluated from the experimental tracking of the velocity of the droplet as it approaches the wall, is on the order of micro-Newtons. The droplet moves at a velocity of centimeters per second. The average velocity depends on the volume of the droplet and the concentration of the solution (Fig. 3(a)). The velocity follows a power law with respect to volume (Fig. 3(a)). For droplets with the same volume, the velocity linearly depends on the concentration of the aqueous solution (Fig. 3(b)). We confirmed that the aniline liquid pseudo-partially wets the water surface. The Marangoni-driven spreading of aniline proceeds through a precursor film and aniline molecules diffuse toward the area with a higher surface tension. During this motion, the contact angles at the front and rear are different, as shown in the side view of the droplet (Fig. 4). The difference in contact angles indicates that the surface tension should be different at the front and rear. When the droplet is static on the saturated solution, contact angle is uniform along the contact line (Fig. 4(b)). Thus, we can estimate the driving force per unit length using the difference in the contact angle, $f_D = \gamma_{o/s}(\cos\beta' - \cos\beta)$, where contact angles $\beta$ and $\beta'$ are illustrated in Fig. 4(c) and $\gamma_{o/s}$ is the surface tension at the oil-solution interface. The power law of velocity with respect to the volume of droplets is predicted using the balance between the driving force and viscous drag. The basic mechanism of the regular motion of aniline droplets is the symmetry breaking of Marangoni-driven spreading. As shown in Fig. 4(d), fluid flow distorts Marangoni flow from the contact line. Marangoni flow takes the aniline molecule, which is surface-active, to the rear of the droplet. the surface tension at the rear is lower than that at the front. The imbalance in surface tension supplies the driving force for the continued motion of the droplet. The system transfers the chemical energy deposited in the bulk liquid to kinetic energy through the continuous spreading of aniline on the air-solution interface. When a droplet approaches near a wall, the wall prevents the further spreading of aniline molecules and reduces the difference in surface tension between the front and rear. Thus, the wall repels the droplet. In a circular Petri dish, a droplet prefers circular motion along the wall of the dish. Centripetal force is supplied by the meniscus effect (deformation of the air-water interface) and the difference in surface tension between sides of the droplet. The essential features of the motion of a self-propelled droplet can be described using a simple differential equation

$$\frac{d\bar{U}'}{dt} = -\xi(|\bar{U}'|^2)\bar{U}' - \nabla\varphi \tag{1}$$

where the velocity-dependent active friction coefficient is $\xi(|\bar{U}'|^2) = \mu(|\bar{U}'|^2 - U^2)$ and $\varphi$, $\bar{U}'$, $|\bar{U}'|$, $U$ and $\mu$ are the effective potential from the repulsion of the wall, velocity, absolute velocity, steady velocity and a constant, respectively. The effective potential has been constructed according the experimental tracking. The numerical results show that the initial condition determines the subsequent mode of motion and the circular mode of motion has been found in the simulation when a droplet has a specific initial velocity. This model reproduces the essential aspects of the spontaneous motion of the droplet [10].

## 3. Dynamic labyrinthine pattern

Figure 5 shows an example of the dynamic labyrinthine pattern in an active liquid film in the dewetting-wetting process [11]. A circular Petri dish is filled with pure water and the surface of water is covered with a layer of pentanol film. A small hole is formed by blowing the surface of pentanol film using a pipette. The hole grows by dewetting. In the initial stage, the pattern emerges following dewetting of a contact line at a constant velocity, which obeys the Culick law [12]. The contact line then develops complex patterns, as shown in Fig. 5(b). After the hole grows, the dynamic labyrinthine pattern exists for a long time (on the order of hours). The morphology of the pattern changes continuously. The contact line forms an interesting labyrinth, where the labyrinthine pattern changes dynamically with time. The dynamics of the patterns is rather different from those of stationary patterns [13, 14]. For a single-domain pattern, the evolution of the pattern exhibits periodic behavior because of interaction with the wall of the Petri dish. Autoregressive behavior is found during the evolution of such a pattern in a Petri dish. The system is far from equilibrium. The contact lines repel each other. When two contact lines approach each other, damping interaction exists between them. Two contact lines do not fuse into each other unless they move toward each other at high velocity. Thus, a contact line can evolve into a complex labyrinthine structure.

The motion of the contact line is driven by a Marangoni-driven spreading process [10]. Two antagonistic interactions, tension in the pentanol film and surface tension at the air-water interface, control the evolution of the contact line. By considering the motion of fluid in the pentanol film and the diffusion of pentanol molecules at the air-water interface, we found that the evolution of the pattern formed by the contact line is controlled by a diffusion process. This is similar to the solidification process, which is controlled by the diffusion of heat in the vicinity of the solidification front [15]. The motion of the interface is driven by the curvature of the moving front. Since the thickness of the pentanol film is on the order of a millimeter, an inertial effect plays a role in the wetting and dewetting processes of the liquid. This inertial effect is one of the reasons for the dynamic behavior of the pattern.

If we consider the evolution of a single-domain pattern for simplicity, the motion of the contact line as a closed curve can be described using the following model [11]:

$$\rho_l \frac{d\upsilon}{dt} = -\zeta\upsilon + \nu\kappa + \Lambda_b (A_0 - A) \tag{2}$$

$$\dot{\kappa} = -[\kappa^2 + \frac{\partial^2}{\partial s^2}]\upsilon \tag{3}$$

where $\upsilon = \dot{\vec{r}} \cdot \hat{n}$ and $\hat{n}$ is a normal unit vector on the contact line, $\zeta$ is a friction coefficient, $\nu$ is a constant, and $\kappa$ is the curvature of the contact line. We parameterize the closed boundary of the pattern as $s \in [0, s_0]$, with generalized coordinates $\vec{r}(s)$ and velocity $\dot{\vec{r}}(s)$, and $s$, $s_0 = s_0(t)$, $\rho_l$, $A$ and $A_0$ are the arc length of the contact line, the total arc length, the density of the contact line, area of the pattern and the equilibrium area of the pattern, respectively. For a closed curve, the dynamics of the area can be described using $\frac{\partial A}{\partial t} = \oint_l ds\upsilon$, where $l$ is the closed curve. This model reproduces the essential aspects of the dynamics behavior of the pattern, as shown in Fig. 6. The morphology of the domain changes dynamically during evolution. In the numerical simulation, the coalescence of contact lines and repulsive interaction between two contact lines are considered. The pattern will split during evolution and the small part is omitted in the result. The pentanol-water system is an active system with self-agitation. When a droplet of pentanol liquid is deposited on the aqueous solution of pentanol, spontaneous motion of the droplet is observed [16]. The chemical nonequilibricity of surface tension in the vicinity of the contact line causes spontaneous motion of the interface. In this pattern evolution, external disturbance due to the instability of Marangoni-driven spreading induces complex dynamics of the pattern. Dissolution and volatilization at the air-water interface decrease the surface density of surface-active molecules and provide the continuous driving force for the Marangoni-driven spreading. Dynamic pattern formation is an important focus of study in the behavior and neural systems [17, 18]. Our results may lead to a better understanding of the mechanism of dynamic pattern evolution in biological systems.

## 4. Conclusion

We have described the self-propelled motion of a droplet and dynamic pattern formation under conditions far from equilibrium from both experiment and theoretical perspectives. As an interesting phenomenon, we found that droplets exhibit repulsive interaction between each other and toward a wall. This suggests that it may be possible to control an individual droplet. These droplets may be suitable for observing swarming phenomenon. When a phase boundary exhibits self-agitation, dynamic pattern formation is observed. A curvature-governed theoretical model has been discussed. The effects of the boundary of the Petri dish and the inertial effect of motion were also considered, and these lead to the dynamic evolution of the pattern morphology. This theoretical model reproduces the essential aspects of dynamic patterns.

Figure captions

FIG. 1 Self-propelled motion of droplets. (a) Spatio-temporal evolution of beeline motion (left) and circular motion (right) in a circular Petri dish. The droplets have volumes of 1800 μl (left) and 500 μl (right), respectively. (b) Spatio-temporal image, top view, of beeline motion between two parallel walls in a square vessel of 10cm×10cm. The concentration of the solution in (a) and (b) is 2.8 vol% and 3.0 vol%, respectively. Scale bars in (a) are 2 cm.

FIG. 2 Experimental tracking of the self-propelled motion of a aniline droplet in a circular Petri dish. (a) A typical trajectory of beeline motion. The volume of the droplet is 1000 μl. (b) Mode-switching from beeline motion to circular motion. The volume of the droplet is 400 μl. Scale bars are 2 cm. The concentration of the solution is 2.8 vol%. The arrows show the direction of motion. (a) is adopted from Ref. [10] with slight modification.

FIG. 3 Velocity of a self-propelled droplet. (a) $\text{Log}_{10}$-$\log_{10}$ plot of the velocity (unit: cm/s) and volume of a droplet (unit: μl) in motion on various solutions (Experiment). The velocity is the average velocity. B: beeline motion in a square vessel (500 ml of solution), C: circular motion in a circular Petri dish (200 ml of solution). The lines in the figures fit to the data and $\eta$ is the slope of the fitting lines. (b) Velocity depending on the concentration of the solution. The volumes of the droplets are 100μl, 150μl 300μl, 600μl, 900μl and 1200μl, respectively. The lines are fit to the data. (a) is adopted from Ref. [10] with slight modification.

FIG. 4 Scenario of a self-propelled droplet at an air-solution interface. (b) Side view of a droplet in beeline motion (left: lateral view, right: front or back view). (b) Side view of a droplet on a saturated solution of aniline. (c) Precursor film model of a droplet from which aniline spreads on an aqueous solution. (d) Illustration of fluid flow (dashed line) and Marangoni flow (solid line) near a droplet in motion. The Reynolds number is from 40 to 200. The fluid flow consists of not only layer flow but also "Karman vortex street" at the rear. The Marangoni flow is distorted by fluid flow. (c) and (d) are adopted from Ref. [10] with slight modification.

FIG. 5 Dynamic labyrinthine patterns. (a) Schematic of the experiment. (b) Typical dynamic labyrinthine pattern. After the initial growth of a hole, a dynamic labyrinthine pattern emerges. The initial thickness of the film is about 1.50 mm. The diameter of the Petri dish is 18.0 cm. Scale bar is 3cm. (a) and (b) are adopted from Ref. [11] with slight modification.

FIG. 6 Numerical results for the time-development of a single-domain pattern. We set $\nu = 1.0$, $\rho_l = 1.0$, $\zeta = 0.5$, $\Delta t = 0.002$, $\Lambda_b = 0.2$ and $A_0 = 20.0$ in the simulation. Splitting of a domain and the coalescence of contact lines are indicated by circles and arrows. The figure is adopted from Ref. [11] with slight modification.

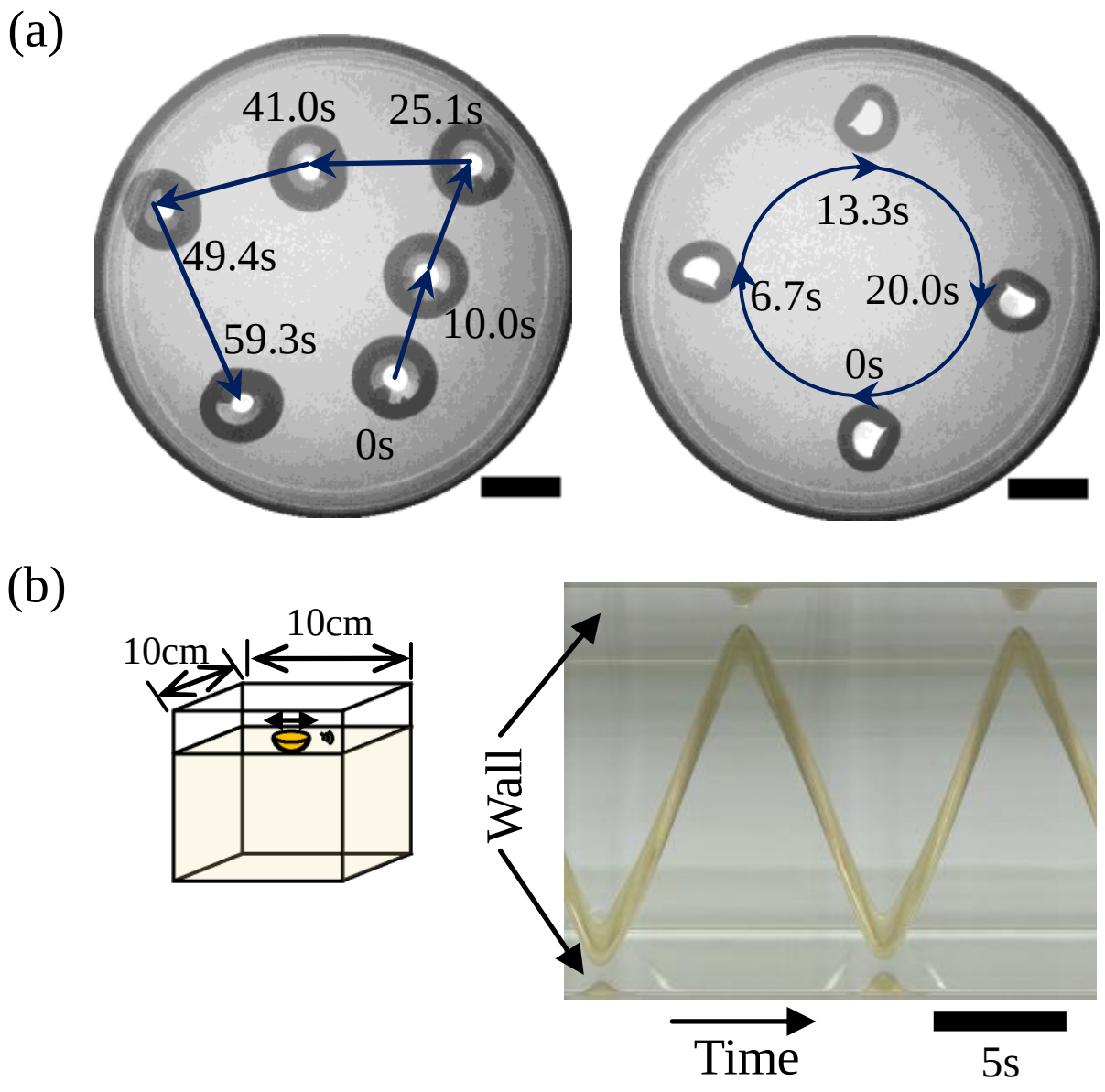

FIG. 1

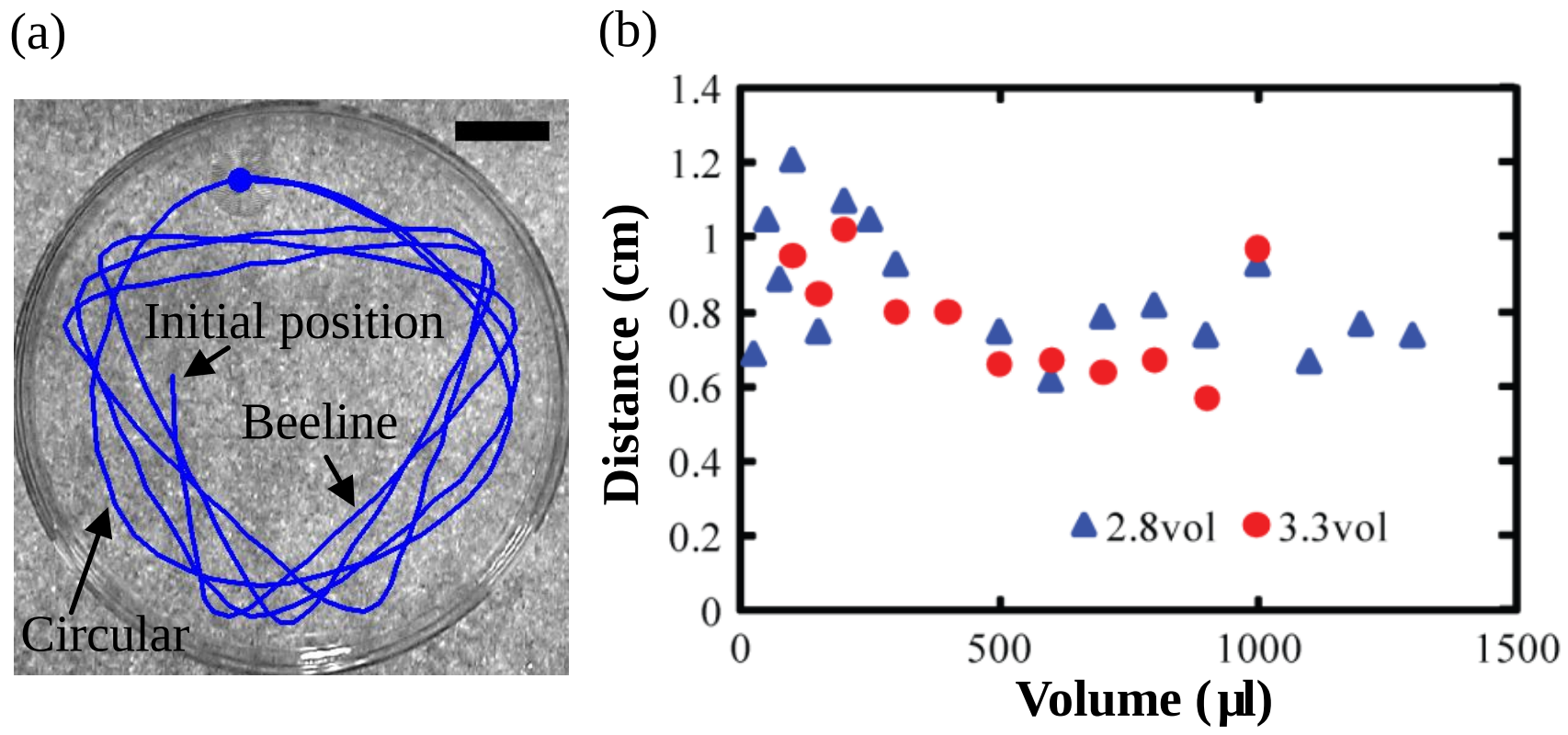

FIG. 2

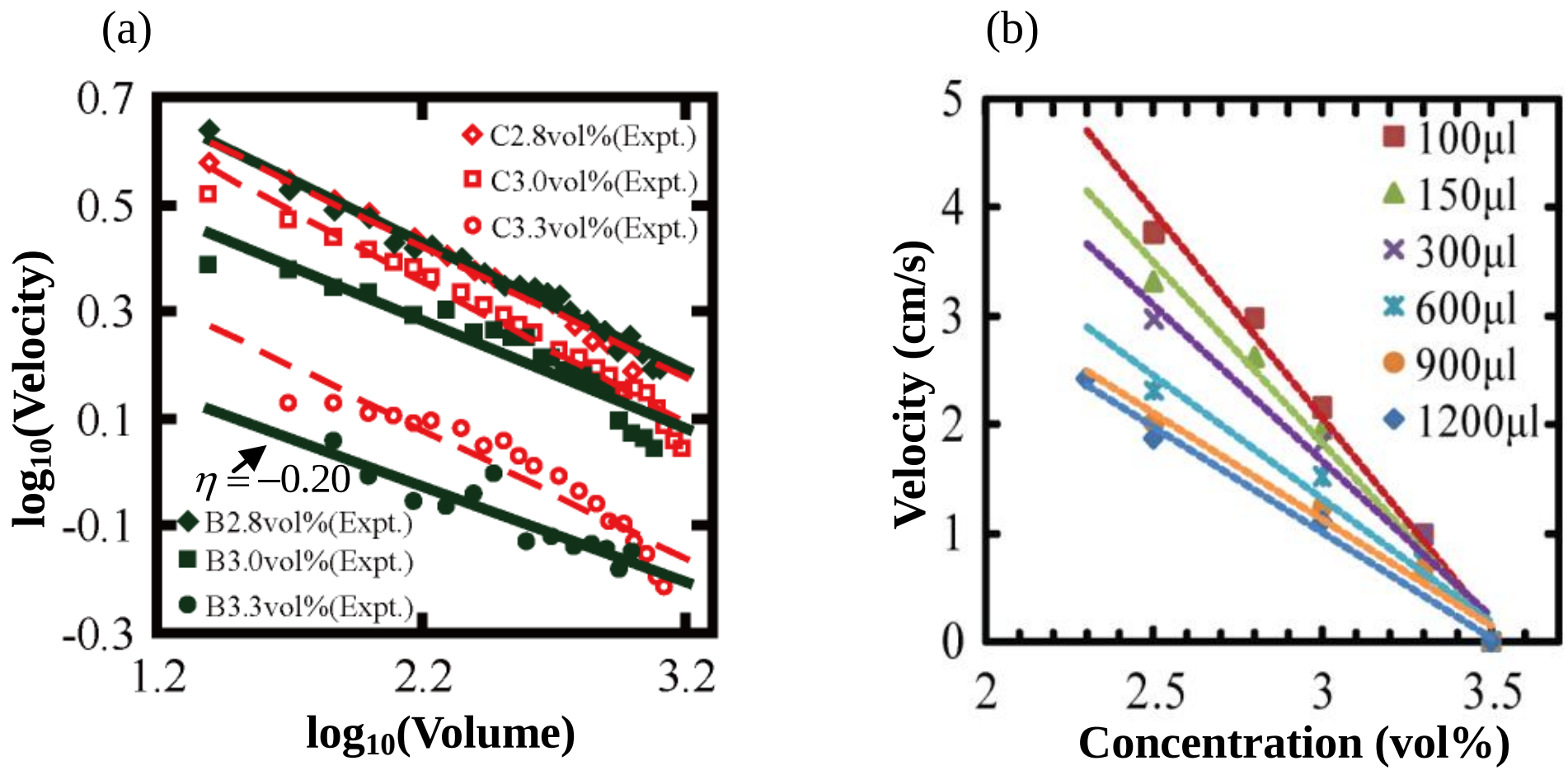

FIG. 3

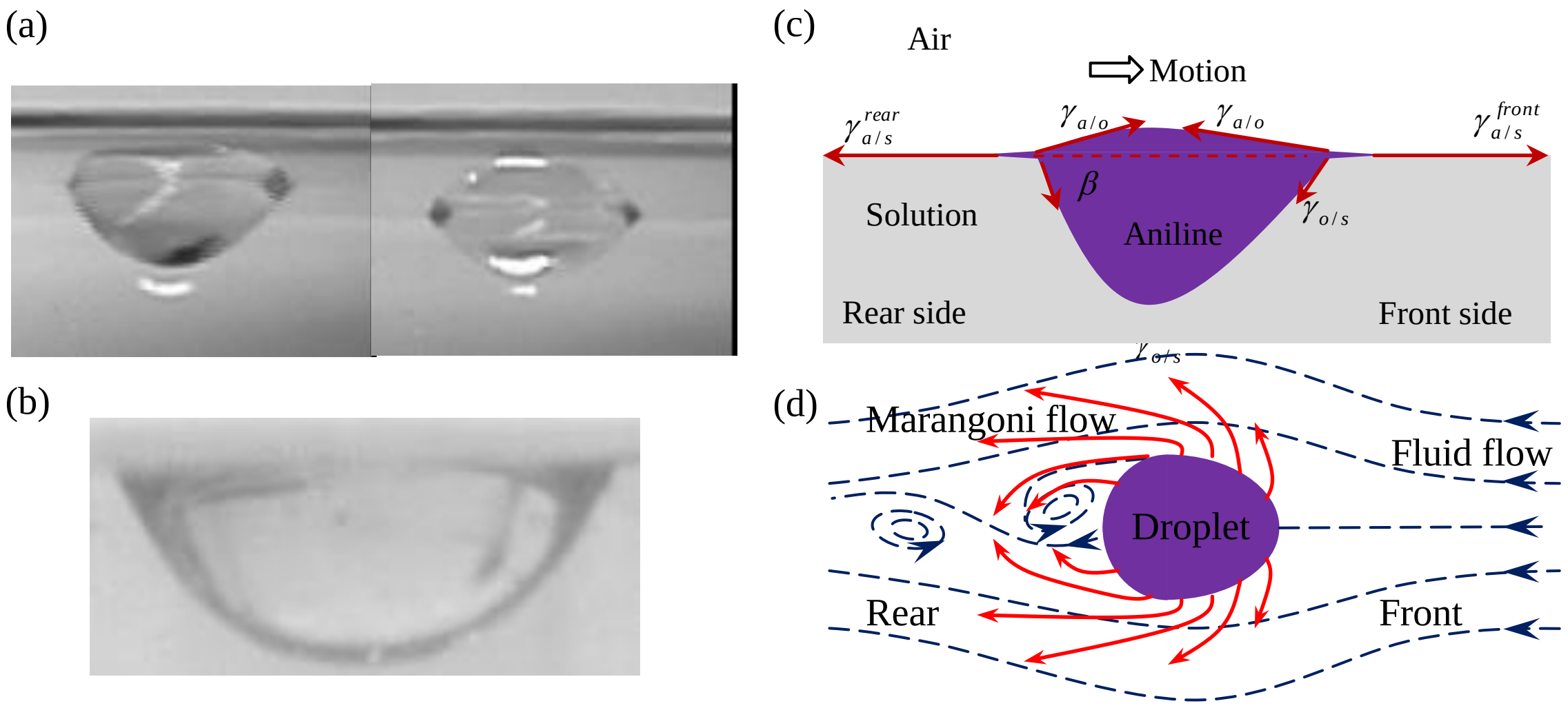

FIG. 4

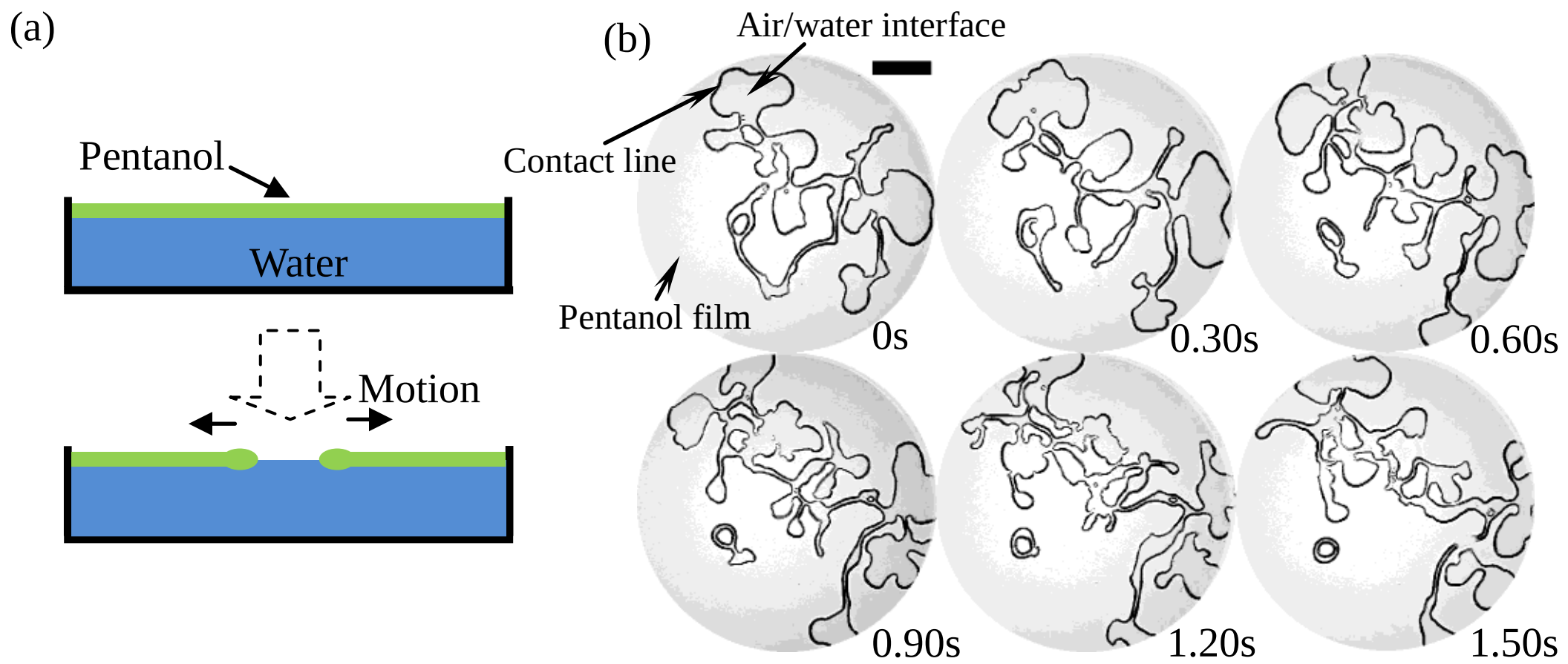

FIG. 5

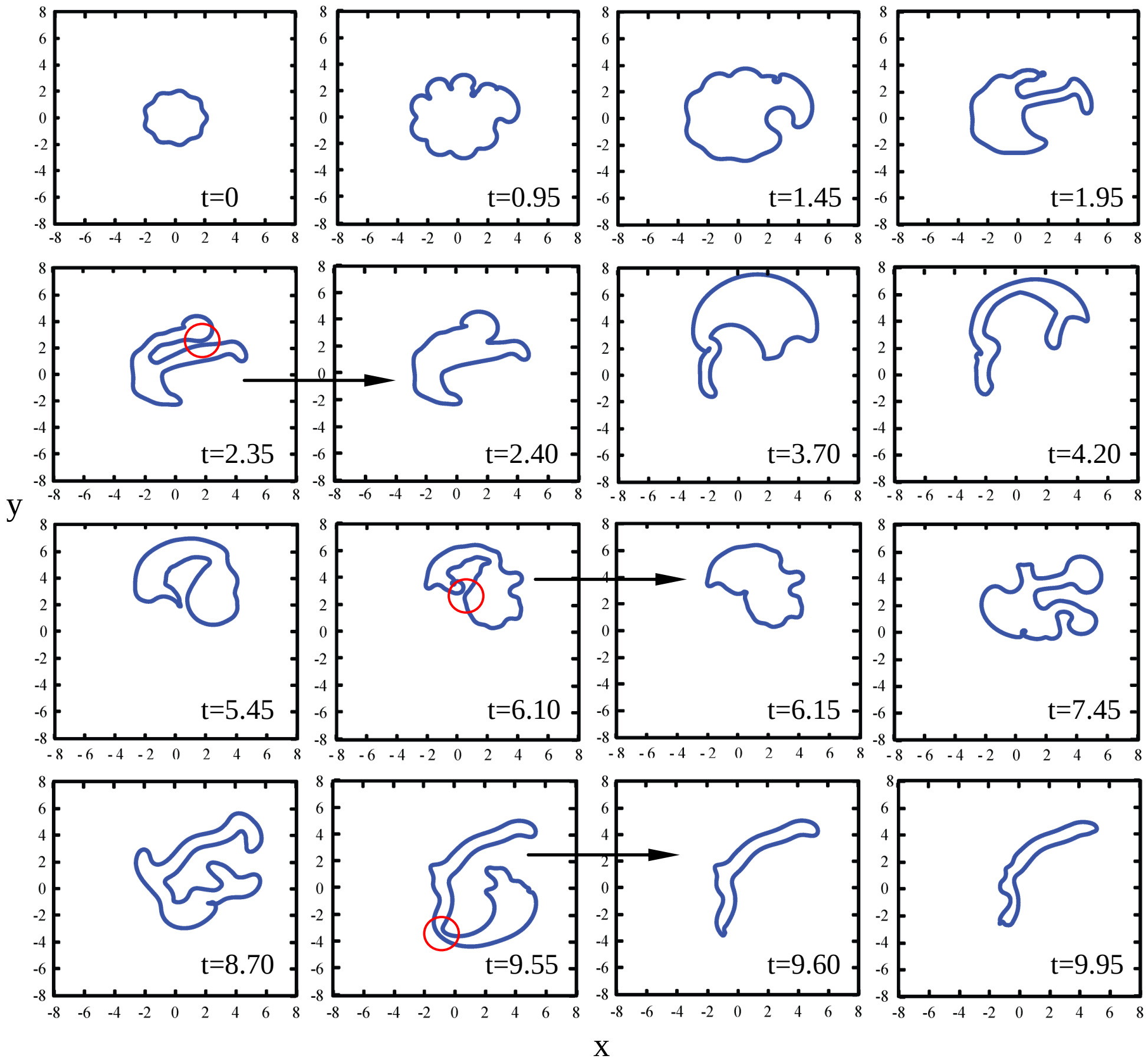

FIG. 6